\documentclass{svmult}

\usepackage{graphicx}        

\begin{document}

\title*{Signal Processing and Control in Nonlinear Nanomechanical Systems} 
\author{R. L. Badzey\inst{1}, G. Zolfagharkhani\inst{1}, S.-B. Shim\inst{1}, A. Gaidarzhy\inst{1}\and
P.~Mohanty\inst{1}}
\institute{Department of Physics, Boston University, Boston, MA 02215, USA}
\authorrunning{R. Badzey {\it et. al.}}
\maketitle

\section{Introduction}
\label{Intro}
Bestriding the realms of classical and quantum mechanics, nanomechanical structures offer great promise for a
huge variety of applications, from computer memory elements \cite{badzey04} and ultra-fast sensors to quantum computing.
Intriguing as these possibilities are, there still remain many important hurdles to overcome before nanomechanical
structures approach anything close to their full potential. With their high surface-to-volume ratios and sub-micron
dimensions, nanomechanical structures are strongly affected by processing irregularities and susceptible to
nonlinear effects. There are several ways of dealing with nonlinearity: exceptional fabrication process control in
order to minimize the onset of nonlinear effects or taking advantage of the interesting and oftentimes counterintuitive
consequences of nonlinearity.

A fundamental control mechanism popular for its counter-intuitive ability to amplify coherent behavior via the
addition of noise, stochastic resonance (SR) has waxed and waned in popularity since its inception over 20 years ago. Originally, it was
postulated \cite{benzi81} as an ad hoc explanation for the periodic onset of ice ages over the Earth's climate history. Given a
nonlinear system with an energy threshold subject to a sub-threshold periodic modulation and white noise, there is
a certain regime of added noise power in which the system oscillates between its states. This oscillation is
synchronized with the modulation. Although this theory has since fallen out of favor in the climate-modeling
community, the appeal of the concept encouraged its extension to a large variety of systems. These include (but
are by no means limited to) bistable ring lasers \cite{mac88}, neurophysiological systems (mechanoreceptors in crayfish \cite{doug93}
and crickets \cite{lev96}), SQUIDs \cite{rou95}, mechanical systems \cite{spano92,chancondmat}, electronic systems \cite{fauve83} such as amplifiers, and
geophysical systems \cite{alley01}. However, there have not been any studies demonstrating the effect of stochastic
resonance in nanomechanical systems. Aside from being simply one more system in which the phenomenon has
been demonstrated, nanoscale systems \cite{lee03} are interesting because of their proximity to the realm of quantum
mechanics. The combination of stochastic resonance and quantum mechanics has been the subject of intense
theoretical activities \cite{wellens00, goychuk99, grif96, lof94} for many years; nanomechanical systems present a fertile ground for the study of a
broad variety of novel phenomena in quantum stochastic resonance. Additionally, the physical realization of such
nonlinear nanomechanical strings offer the possibility of studying a whole class of phase transition phenomena,
particularly those modeled by a Landau-Ginzburg quantum string \cite{benzi85, hu99} in cosmology.

\section{Nonlinearity in Nanomechanical Structures}
\label{nanomech}

The nanomechanical structure under consideration is a simple suspended beam, clamped at both ends and
subjected to a transverse driving force. Following continuum mechanics, it is well known that such a structure
has a resonance frequency for the fundamental flexural mode which is given by

\begin{equation}
\label{fvsl}
f\sim\sqrt{\frac{E}{\rho}}\frac{t}{L^{2}}\, 
\end{equation}
where E is the Young's modulus, $\rho$ is the material density, t is the thickness (dimension parallel to the transverse forcing), and L is the length of the beam. Our doubly-clamped beams were fabricated from silicon, with
dimensions of 7-8 $\mu$m x 300 nm x 200 nm; these yielded resonance frequencies around 25 MHz. Clamping losses
and electrode mass loading reduced the resonance frequency from that which could be assumed from the simple
continuum mechanics derivation. Fabrication of these structures is a well-established sequence of techniques,
including PMMA spinning, e-beam exposure, development, metallization, and both dry and wet etching
techniques. The details of fabrication are given in several locations \cite{badzey04, badzey05}.
The magnetomotive technique has been in use for many decades, providing an efficient and low-noise
method for exciting the resonant modes of a suspended structure. A sinusoidal current is pushed through a
metallic electrode in the presence of a magnetic field. The resulting Lorentz force is then:

\begin{equation}
\label{lorentz}
\overrightarrow{F}\cos\omega t=\overrightarrow{I}l\cos\omega t\times\overrightarrow{B}\
\end{equation}
It is easy to see that the Lorentz force will be perpendicular to both the field and the current; a field perpendicular
to the plane of the substrate will produce an in-plane transverse vibration. This in turn produces a motional EMF
on the two clamping electrodes. This voltage is proportional to the velocity of the beam:

\begin{equation}
\label{magneto}
V_{avg}(t)=\xi lB\frac{dx(t)}{dt},\
\end{equation}
where the factor $\xi$ is a proportionality constant given by the mode shape. For a fundamental mode, $\xi$ = 0.53. In
the harmonic approximation, this induced voltage is directly proportional to the displacement of the beam. Under
this force, the beam will behave as a damped, driven harmonic oscillator:

\begin{equation}
\label{linear}
m\ddot{x}+\gamma\dot{x}+kx=F\cos\omega t.\
\end{equation}
The amplitude of motion describes a Lorentzian lineshape as a function of frequency, centered at the resonance
frequency and with a quality factor Q = $\omega_{0}/2\gamma$. In the limit of small dissipation Q = $\omega_0/\Delta\omega$.
There are two main roads into the nonlinear regime: inherent nonlinearities in the suspended bridge, or
increasing the driving force until the beam response becomes nonlinear. Both methods result in the equation of
motion described by the Duffing equation,

\begin{equation}
\label{duffing}
m\ddot{x}+\gamma\dot{x}+kx\pm k_{3}x^{3}=F\cos\omega t,\
\end{equation}

\noindent
where the addition of the cubic restoring force term yields a dramatic change in the resonance behavior. It is a
well-known phenomenon that a Duffing oscillator exhibits hysteresis and bistability near the linear resonance
frequency. The evolution of an oscillator from linear response into nonlinear behavior is well-studied and
covered by a number of excellent texts – a simple qualitative explanation will suffice. Simply put, as the beam is
subjected to ever-increasing driving forces, it stretches in response, thereby increasing its length. The clamping
points being a fixed distance apart, the situation is reached whereupon the length of the bridge exceeds that
between the clamping points. Therefore, when the bridge passes through equilibrium, it feels a compressive force
from the pads, giving rise to a quartic term in the beam potential and therefore a cubic term in the equation of
motion. This is the well-known Euler instability, which occurs under the influence of any compressive or tensile force. For a doubly-clamped beam, there is a critical force which describes the boundary  between the linear and nonlinear regimes \cite{feynmann}:

\begin{equation}
\label{critforce}
F_{c}\sim\sqrt{\frac{\gamma^{3}}{k_{3}\omega_{0}}}.\
\end{equation}

\noindent
Depending on the elasticity of the material, it is possible to switch between the linear and nonlinear response
regimes at will by simply tuning the driving force. Eventually, however, the elastic response gives way to a
plastic deformation of the bridge and the nonlinear effects become more and more prevalent and permanent.
When excited nonlinearly, the response of the bridge evolves from the simple Lorentzian into curved peak with a
dramatic drop as seen in the following figure.

\begin{figure}
\centering
\includegraphics[width=2in]{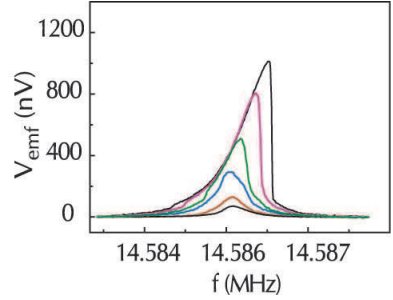}
\caption{Beam response with increasing drive force, showing the transition from the linear to nonlinear regime.  The labeled curves represent the linear response regime of the oscillator in response to different applied forces, from 0.5 pN (curve A) to 3 pN (curve D).  Beyond this critical force, the beam response is nonlinear and bistable.}
\label{powergraph}
\end{figure}

Once the beam is driven into a nonlinear regime it demonstrates hysteresis as a function of frequency, as
predicted by mechanics. As the frequency is increased, the amplitude of the response function follows the
pseudo-Lorentzian lineshape until it reaches the sharp drop as seen in Figure 1. In reality, the full analytic
solution of the frequency-response function shows a rather dramatic bend to lower frequencies, creating a frequency domain in which the amplitude function is multi-valued. Therefore sweeps in one direction in
frequency follow the upper curve of this region, while opposite sweeps follow the lower curve. Eventually, each
sweep reaches a frequency in which the system is unstable and therefore causes a transition to the other state.
This is clearly illustrated in Figure 2.

\begin{figure}
\centering
\includegraphics[width=2in]{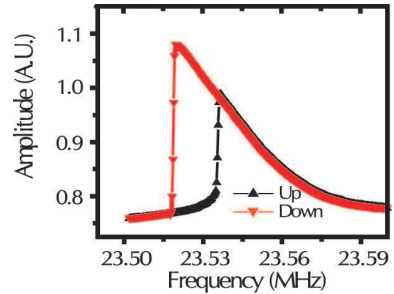}
\caption{Hysteresis caused by the nonlinear bistable response of a silicon nanomechanical bridge under strong
driving.}
\label{hysteresis}
\end{figure}
As stated previously, the Euler instability occurs under the influence of either a compressive or tensile strain on
the beam. These two situations lead to qualitatively different but dynamically similar situations. In the Duffing equation, this is taken into account by the alternative sign on the cubic term,

\begin{figure}
\centering
\includegraphics[width=3in]{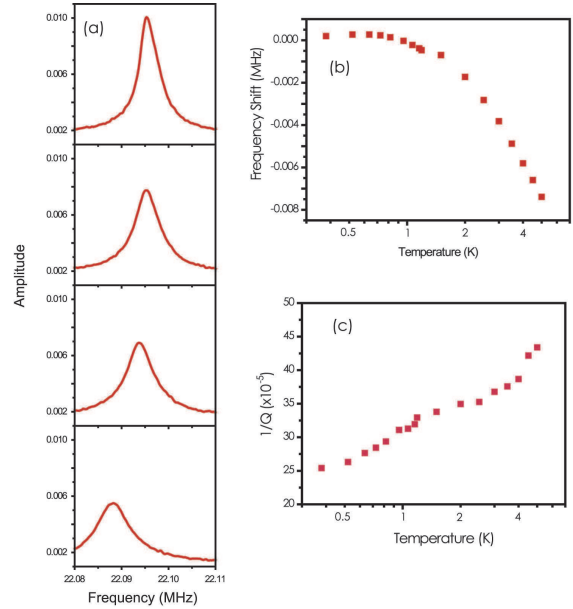}
\caption{Temperature effects on the dissipation and frequency shift of a nanomechanical silicon bridge. (a). Variations
in the resonance peak with temperature. As the temperature is increased the peak gets broader and also undergoes
shift to lower frequencies. (b). Graph of the frequency shift as a function of temperature. The total shift is
approximately 9.5 kHz, or $4 \times 10^{-4}$ of the initial resonance at 300 mK. (c). The dissipation (inverse Q) as a function of
temperature. This quantity changes by almost a factor of 2 over the temperature sweep.}
\label{dissvst}
\end{figure}

\noindent
with compression yielding a positive
sign and tension a negative sign. These two situations also present slightly different frequency response spectra
as well. The compressive case is revealed by a sharp drop on the right-hand side; the tensile spring has a drop on
the left. For illustrative purposes, Figure 1 shows a compressive case, while Figure 2 shows what appears to be a
tensile case. Regardless of whether there is compression or tension, the oscillator is still bistable and
hysteretic.
The fact that bistability arises in both nonlinear cases is important, because the
effect the electrical measurement circuit can have on the apparent behavior of the oscillator is worth noting. It is understood that
a mechanical harmonic oscillator system can be written as an equivalent LCR circuit \cite{mohanty02}– by the same token an
electrical system can be constructed as a mechanical one. Therefore, when a mechanical system is connected to
an electrical LCR measurement circuit, the signals measured are due to both constituent parts. The electrical
resistance, capacitance, and inductance can be related to the mechanical spring constant and dissipation

\begin{equation}
\label{mechelec}
R\sim\frac{1}{\gamma k},\quad C\sim k\quad L\sim\frac{1}{k}.\
\end{equation}

\noindent
These relationships also hold as the oscillator evolves into nonlinear response -- the electric circuit need not be nonlinear for its signal to combine with that of the oscillator. Therefore, the
relationship between the effective linear ($k$) and nonlinear ($k_{3}$) spring constants of the combined electromechanical
system can be either compressive (same sign) or tensile (opposite sign). Regardless, the physical state
of the mechanical structure itself is determined by the physical geometry of the system – in all of the cases
described here, the compressive response is the physically realistic one even though the frequency response might
suggest otherwise. This supposition was borne out by an experimental check -- changing the cables (to ones with
different electrical parameters) resulted in a flip of the nonlinear response from a left-hand drop to a right-hand drop. 

\begin{figure}
\centering
\includegraphics[width=2in]{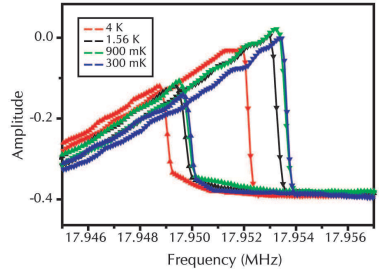}
\caption{Alteration of the hysteresis region by an increase in the temperature from 300 mK to 4 K.}
\label{hystvst}
\end{figure}

In addition, since both tensile and compressive cases result in bistability, the distinction from the standpoint
of device applications is a largely academic one.
From the derivation of the frequency response, assuming that the solution can be written as a harmonic
function, the states are of the vibrational amplitude. In actuality, the most accurate statement is that they are
different vibrational velocity states. This is borne out by the response seen by the magnetomotive technique,
which is at its heart a velocity measurement. Only should the harmonic approximation prove valid can it be said
that the two states are indeed amplitude states.
For all of the illustrative abilities of the previous analyses, they all neglect the effect of temperature on the
oscillatory characteristics of the bridge. At room temperature, slight fluctuations will have a negligible effect, and
large variations in temperature are neither realistic nor easy to implement. Low temperatures, however, are easily
attained through a variety of techniques. We inserted a nanomechanical silicon bridge into the cryostat of a $^{3}$He
refrigerator and varied the temperature from 300~mK up to 5~K. 

As is clear from the Figure 3, both the spring
constant and the dissipation were affected by changes in temperature. The first panel is a series of frequency
scans taken at a variety of temperatures. Changes in temperature have two separate but related effects. First is a
shift in the resonance frequency largely due to a reduction of the spring constant. The second is a gradual
broadening of the peak and decrease in the amplitude at resonance, stemming from an increase in the dissipation
factor. However, these two results are not completely independent. The frequency shift is also partly responsible
for the change in measured Q, making a true separation of the effect of temperature on $\gamma$ and $k$ difficult. However,
to a large degree, measuring Q gives a good approximation of $\gamma$, and measuring the shift in frequency sheds light
on the change in $k$. 

From 300~mK to 5~K, the frequency shifts by 9.5~kHz; taken as a fraction of the initial
resonance frequency of approximately 22.0945~MHz, this shift is $4 \times 10^{-4}$. This means that the spring constant
also changes by approximately four parts in $10^{4}$, as $\Delta f\sim\Delta k$. Interestingly, what is also clear from this temperature
sweep is the effect on the dissipation, measured by tracking the changes in the Q of the oscillator. While the
frequency shift is negligible, the change in dissipation most certainly is not. The dissipation (1/Q) at 300~mK has
a value of $25 \times 10^{-5}$; heating to 5~K increases this value to $45 \times 10^{-5}$, an increase by almost a factor of two.

It is important to note that the preceding data comes from a nanomechanical oscillator in linear response. In
nonlinear response, the notions of resonance frequency and Q become very difficult to formulate. However, the
characteristics of the hysteresis region itself contain information about the behavior of $\gamma$, $k$, and $k_{3}$. These three parameters are even more closely linked than in the linear regime. As the concept of resonance frequency is somewhat misleading for the nonlinear case, the better measure is the central frequency of the nonlinear hysteresis region. By the same token, the concept of Q as the ratio between the resonance and the FWHM of the peak loses merit in the nonlinear case; a measurement of the change in the width of the hysteresis is a better indicator. Again, extracting the exact contribution due to each particular parameter is difficult when looking at the temperature alone.
When the temperature is increased as shown in Figure 4, there are two main results – a slight shift to lower
frequencies and a reduction in the width of the hysteresis. From a low of 300~mK to a high of 4~K, the central
frequency of the hysteresis region shifts by approximately 1.25~kHz, and the width decreases by approximately
500~Hz. It is important to note, however, that transitions within the hysteresis region are probabilistic and
affected by the ambient temperature. Many sweeps through the hysteresis region are required in order to fully
characterize the frequency shift and width change. In general, though, this result agrees with earlier experiments
\cite{cleland05} that examined the change in the nonlinear characteristics with the addition of white noise.

\section{Control by Stochastic Resonance}

One of the more interesting nonlinear effects to come forth in the past few decades is the phenomenon of
stochastic resonance. Compelling because of its ability to draw out coherence from a background of noise,
stochastic resonance has suffered a bit in recent years because of a dearth of new and interesting experimental
observations. Most have been received as simply one more example of stochastic resonance in a system, with
little impact beyond adding to a pile of curiosities. However, the presence of nonlinear behavior in a
nanomechanical system \cite{badzeynat} opens up an interesting realm of inquiry, a new avenue into the exploration of the
phenomenon. 

These nanomechanical systems, by virtue of their small sizes, high frequencies, and low
temperatures venture quite close to the regime of quantum mechanics. Additionally, these nanomechanical
bridges are closely related to a system which has become even closer to reality – a truly macroscopic, mechanical
quantum harmonic oscillator \cite{quantprl, quantapl}. One of the very basic and necessary conditions for any oscillator to exhibit
quantum behavior is that its modal energy must be greater than the thermal spreading of the energy states. That
is, $hf > k_{B}T$, where $k_{B}$ is the Boltzmann constant and $h$ is Planck's constant. For an oscillator with a resonance
frequency of 1~GHz, this condition is achieved when the temperature is below 48~mK. Both of these are
experimentally realizable conditions, and several interesting experiments have been put forth which show the
deviation of such oscillators from the classical norm. Therefore, exhibiting stochastic resonance in these MHz-range
nanomechanical oscillators can lead to exploring the phenomenon in the closely related oscillators beyond
the quantum limit. At the quantum level, prying coherent behavior out of a noisy environment could prove to be a powerful means of achieving quantum control. 

\begin{figure}
\centering
\includegraphics[width=3.5in]{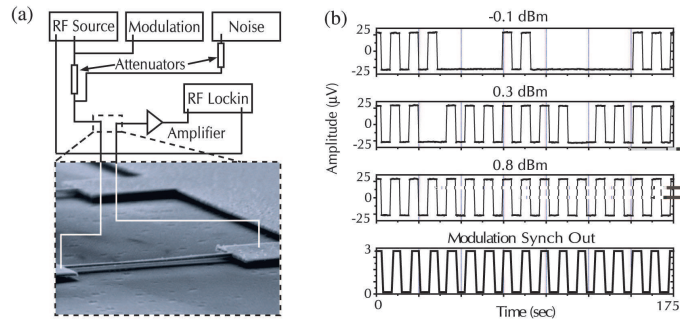}
\caption{(a) Schematic of the electrical circuit used to excite a nanomechanical silicon oscillator into nonlinear response, modulate switching between the two resultant bistable states, and introduce electrical broadband noise. (b) Sample switching events, showing the effect of reducing modulation. The bottom panel shows the synch out from the modulation source.}
\label{circuit}
\end{figure}

To look more universally, these systems closely approximate
quantum strings described by a Landau-Ginzburg equation.

\begin{equation}
\label{landauginzburg}
\frac{\partial\Phi(x,t)}{\partial t}=m\Phi(x,t)-\Phi^{3}(x,t)+\kappa\frac{\partial^{2}\Phi(x,t)}{\partial t^{2}}+A\cos(\Omega t)+\eta(x,t)\
\end{equation}

\noindent
Here, the field variable $\Phi$ replaces the position variable $x$, and $\eta$ is the noise term. Therefore, these
and related nanomechanical structures provide a powerful tool for examining fundamental properties of a whole
host of systems which are governed by similar equations and the resultant phase transition effects.
The central requirement for the observation of stochastic resonance is a nonlinear two-state system subjected to
a subthreshold modulation in the presence of tunable white noise. With the well-behaved bistable behavior seen
in these nanomechanical oscillators, they present a very clean system in which to study the effect. Also, as the
temperature and the driving power affect the hysteresis, the beam characteristics can be adjusted so as to
study an entire parameter space.

For this experiment, we used two silicon bridges, cooling them in the $^3$He cryostat and exciting in-plane
vibrational modes with the magnetomotive technique. At 300~mK bridge 1 had a linear resonance frequency of
approximately 23.57~MHz (Q = 3700), while bridge 2 had a linear resonance frequency of 20.835~MHz (Q =
1000). Both exhibited nonlinear bistable behavior under the influence of suitably strong forcing (4.0~dBm for
bridge 1, 1.0~dBm for bridge 2). It was found during previous studies of bridge 1 that it was possible to excite the
oscillator nonlinearly at a single frequency within the bistable region. With the addition of a square-wave
modulation it was then possible to control the state of the oscillator, forcing it back and forth between its two
states with the modulation frequency $\Omega$.  

Under the influence of the driving force and modulation, then, the
equation of motion for either of these bridges is a modified Duffing equation of the form:

\begin{equation}
\label{array}
\begin{array}{c}
m\ddot{x}+\gamma\dot{x}+kx\pm k_{3}x^{3}=F_{d}\cos\omega_{d}t+f_{m}(\Omega)\\
f_{m}(\Omega)=\frac{F_{m}\Theta(t)}{2};\quad\Theta(t)=\left\{ \begin{array}{c}
-1\quad(n-1)T<n<n-1/2)T\\
1\quad(n-1/2)T<t<nT\end{array}\right.\end{array}\
\end{equation}

\noindent
This shows the equation with square wave forcing, although in principle any modulation term can be used. The
response of the bridge was highly dependent on the amplitude of the modulation, with low powers leading to a
loss of switch fidelity but not in the amplitude of the switches. This is consistent with expectations, as the voltage
difference of the hysteresis is fixed by the driving amplitude and temperature – the modulation only serves to
overcome (or fail to overcome) the existing barrier between the two states. Figure 5 illustrates both the circuit
and sample layout and an example of the switching behavior as a function of modulation power.
Once controllable switching was established, the modulation was reduced to sub-threshold, and a
broadband electrical noise source was introduced into the circuit, and the noise power was swept from –71~dBm
($\sim$ 79~pW) to –41~dBm ($\sim$ 79~nW). At each noise power, a time-resolved scan was performed in order to establish
whether or not the addition of the noise would result in the re-emergence of switching.
Low noise powers resulted in no response from the bridge – it remained in the initially prepared state. As
the noise was increased, however, first sporadic, and then more synchronized switch events were seen. Each
switch event was synchronized with the modulation, and occupied exactly one-half period of the modulation
frequency of 0.05 Hz. If a switch was skipped, no new switch would appear until the modulation passed through
at least one-half period and again entered the positive-going part of the signal. The initial increase in switching
was dramatic and quite rapid, establishing full switching within only a few fractions of a dBm increase in noise
power.
The signature of stochastic resonance is an increase in the signal-to-noise ratio (SNR) as a function of
applied noise power. SNR is defined as:

\begin{equation}
\label{snr}
SNR=\frac{S(\Omega)}{N(\Omega)}\
\end{equation}

\noindent
here, $S(\Omega)$ is the height of the peak in the power spectral density at the modulation frequency, and $N(\Omega$) is the background noise. With the SNR data accumulated, the following graph was compiled, showing a quite dramatic
increase in the SNR with increasing noise power.
An interesting feature of this data is the fact that the SNR does not exhibit the typical sharp peak as a
function of the noise, but is instead rather broad, extending for several dBm before declining sharply down to near
zero again. It should be noted that our system was distinct for several reasons. One, it was subjected to a square
wave modulation as opposed to the canonical sine wave. The effect of a square wave on stochastic resonance has been the subject of considerable theoretical interest \cite{morillo, dykman}. Secondly, the electronic noise source was of broad but
still finite bandwidth -- 15~MHz, to be exact. Canonical theory quite explicitly states that the noise introduced is
Gaussian and broadband. 

\begin{figure}
\centering
\includegraphics[width=4in]{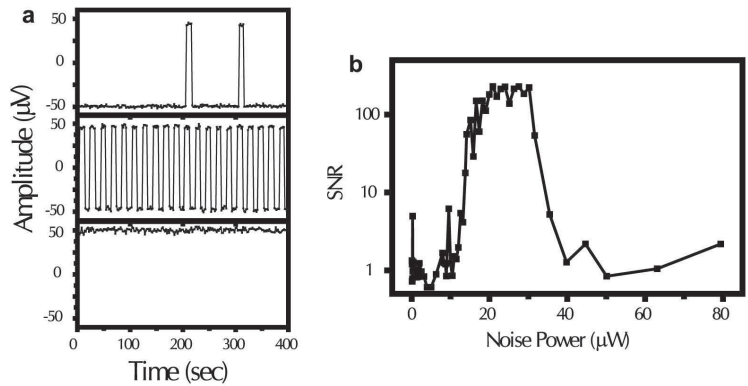}
\caption{SNR as a function of applied noise power. (a) Three representative time series, showing the reemergence
of switching as the noise power is increased. The applied noise increases from the uppermost
panel downward. (b) Taking a power spectrum of each time series yields the SNR for each noise power – these
were combined to create a graph of the SNR versus the applied noise power. The signature increase of
stochastic resonance is clear, even though the overall shape does not follow the expected canonical model.}
\label{SRvnoise}
\end{figure}

Additionally, because of the very clean nature of the system switching,
the barrier between the two states was quite strong. There is a considerable amount of rigidity to the system -- it is
difficult to induce switching between the two and the switching, once begun, is robust. By the same token,
dynamic changes to the conditions of the oscillator were not felt instantaneously. It is possible that a longer time
series at each noise power would reveal subtle differences in the switching behavior that would lead to changes in
the SNR. Finally, the nature of the oscillator also lends itself unusual switching characteristics once the noise has
increased above the effective power for stochastic resonance. In the canonical formulation and most experiments
in stochastic resonance, the decrease in the SNR at higher noise powers is due to a swamping of the switching
signal by a large number of incoherent switches. However, this is not the case here – the oscillator simply stops
switching and instead remains in the upper state. As this is a symmetric system, there is no preference for the
eventual settling in the upper state -- subsequent runs would leave the system in the lower state with just as much
probability. And lastly, the effect of the fact that the noise source bandwidth did not intersect or overlap with the
system resonance frequency should be explored in more detail, as it could contribute to the rigidity. This matter
of system rigidity is certainly an open question, and would benefit from additional experimentation and
theoretical consideration.

As temperature has already proven itself to be a powerful noise source \cite{badzey05} in the behavior of
nanomechanical oscillators, a natural check was to see if stochastic resonance occurred with the source of noise being temperature, not external and electrical. 

\begin{figure}
\centering
\includegraphics[width=4in]{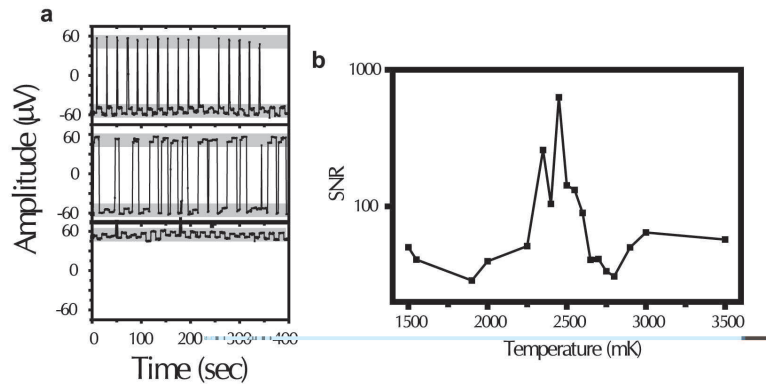}
\caption{SNR as a function of ambient temperature. (a) A sequence of time series for this asymmetric bridge
shows that even when the coherence is regained, it is not as clean as for the first bridge. Also, the first panel,
which shows the sharp spikes and immediate decays, demonstrates the asymmetric character of this bridge. (b)
The SNR graph demonstrates a strong resonance. The data is simply line-connected – there is no fit implied,
and is affected by the coarseness of the sweep.}
\label{SRvTemp}
\end{figure}

The second oscillator was disconnected from the noise source and
brought into nonlinear response. The modulation required for switching was exceptionally strong (19 dBm), to
the extent that the electrical signal appeared in the time series on top of the normally-static beam response
voltage. Again reducing the modulation to below threshold (16.5~dBm), the temperature was increased from 300~mK to 4~K and the SNR was extracted from the time series of the beam response. It is clear from Figure 7, there
is another peak in the SNR as the temperature increases beyond 2~K.
Although it is tempting to apply the same analysis to the SNR response of the second bridge as the first,
the results should be scrutinized carefully. It is obvious from the time series and the power spectra that this bridge
had a non-negligible asymmetry. The presence of asymmetry has been broached in a number of theoretical
investigations \cite{bulsara}, and it can be used to understand this system, as well. 

Because the temperature is changing, its effect on the dissipation and spring constant of the oscillator should not
be overlooked. As shown in the first section, the dissipation is quite strongly changed as a function of the
temperature.  Therefore it is necessary to look at the equation of motion with new eyes. Typically, the
noise considered in the formulation of the stochastic resonance problem is purely additive. However, when the dissipation is a function of temperature, the damping
coefficient $\gamma$ can change the dynamics of the situation. In this case, it would be best to consider the noise as being
partially additive (from the direct effect of heating brought on by the increase in temperature) and partially
multiplicative (from the changes to dissipation). Over the temperature range where the stochastic resonance is
seen (2~K to 3~K), the dissipation (based on the linear data) changes by approximately 10 per cent -- certainly it is an
effect worthy of further investigation. The effect on the spring constant, however, is not large.

In conclusion, we have introduced stochastic resonance into an entirely new class of systems, ones which
by virtue of their small size and high frequencies touch upon an area where this and other nonlinear phenomena
may have a deep and lasting impact. Nanoscale systems, and nanomechanical systems in particular, are much
more than another example of stochastic resonance. Their highly controllable nature makes them perfect
candidates for the study of a host of exciting phenomena, to an extent rarely seen before.

\section{Acknowledgements}
The authors would like to thank M. Grifoni, A. Bulsara, and M. Dykman for their many insightful comments and
thoughtful discussions during the course of these experiments. This work has been supported by the Nanoscale
Exploratory Research grant (No. ECS-0404206) of the National Science Foundation (NSF) and the DOD/ARL (No. DAAD 19-00-2-0004).  This work was also supported in part by NSF grants DMR-0449670 and CCF-0432089.

\end{document}